\documentstyle[mprocl]{article}

\def\be{\begin{equation}}
\def\ee{\end{equation}}
\def\bea{\begin{eqnarray}}
\def\eea{\end{eqnarray}}

\begin{document}
\title{ENERGY CONDITIONS AND GALAXY FORMATION}
\author{MATT VISSER}
\address{Physics Department, Washington University, \\
Saint Louis, Missouri 63130--4899, USA}

\maketitle
\abstracts{
The energy conditions of Einstein gravity are designed to extract as
much information as possible from classical general relativity without
specifying a particular equation of state.  This is particularly
useful in a cosmological setting, where the equation of state for the
cosmological fluid is extremely uncertain. The {\em strong energy
condition} (SEC) provides a simple and robust bound on the behaviour
of the look-back time as a function of red-shift. Observation suggests
that the SEC may be violated sometime between the epoch of galaxy
formation and the present.
}
 
\section{Introduction}
\def\implies{\Rightarrow}
\def\sign{\hbox{sign}}

The energy conditions of Einstein gravity are designed to side-step,
as much as possible, the need to pin down a particular equation of
state. Observation seems to indicate that the SEC is violated rather
late in the life of the universe---somewhere between galaxy formation
and the present time, in an epoch where the cosmological temperature
never exceeds 60 Kelvin.~\cite{Galaxy,Zed15} I show this by using the
SEC to develop a simple and robust bound for the look-back
time as a function of red-shift in Friedmann--Robertson--Walker (FRW)
cosmologies. The experimental observations I need are the present day
value of the Hubble parameter $H_0$, an age estimate for the age of
the oldest stars in the galactic halo, and an estimate for the
red-shift at which these oldest stars formed. From the theoretical
side, I only need to use a FRW cosmology subject to the Einstein
equations and SEC, and nothing more.

\section{SEC in a FRW Universe}

The standard FRW cosmology is described by the metric
\begin{equation}
ds^2 = -dt^2 + a(t)^2 
\left[ {dr^2\over1-k r^2} + r^2(d\theta^2+\sin^2\theta \, d\phi^2)\right].
\end{equation}
with $k = -1,0,+1$.
The Einstein equations are
\begin{equation}
\rho = {3\over8\pi G}
\left[{\dot a^2\over a^2} + {k\over a^2} \right]; 
\qquad \qquad
p = -{1\over8\pi G}
\left[2{\ddot a\over a} + {\dot a^2\over a^2} + {k\over a^2} \right].
\end{equation}
For the case of a FRW spacetime the SEC specializes to

\begin{equation}
\hbox{SEC} \iff \quad 
(\rho + 3 p \geq 0 ) \hbox{ and } (\rho + p \geq 0).
\end{equation}
Consider the quantity

\begin{equation}
\rho + 3 p = -{3\over4\pi G} \; \left[{\ddot a\over a} \right].
\end{equation}
Thus

\begin{equation}
\hbox{SEC} \implies \qquad \ddot a < 0.
\end{equation}
The SEC implies that the expansion of the universe is decelerating---and
this conclusion holds independent of whether the universe is open,
flat, or closed.

\section{Look-back time}

The look-back time, $\tau = |t-t_0|$, is the difference between the
age of the universe when a particular light ray was emitted and the
age of the universe now. 

\begin{equation}
\tau(a;a_0) = |t-t_0| = \int_a^{a_0} {da\over \dot a(a)}.
\end{equation}
By putting a lower bound on $\dot a$ we deduce an upper bound on
look-back time. Since the SEC implies that the expansion is
decelerating, we see

\begin{equation}
\hbox{SEC} \implies \quad 
\tau(a;a_0) = 
|t-t_0| \leq 
{1\over H_0} \; {a_0-a \over a_0} =
{1\over H_0} \; {z\over1+z}.
\end{equation}
This provides us with a robust upper bound on the Hubble parameter

\begin{equation}
\hbox{SEC} \implies \quad 
\forall z: H_0 \leq {1\over \tau(z)} {z\over1+z}.
\end{equation}
This is enough to illustrate the ``age-of-the-universe''
problem. Suppose we have some class of standard candles whose age of
formation, $\tau_f$, we can by some means estimate. Suppose further
that we look out far enough can see some of these standard candles
forming at red-shift $z_f$. Then

\begin{equation}
\hbox{SEC} \implies \quad H_0 \leq {1\over \tau_f} {z_f\over1+z_f}.
\end{equation}

\section{Observations}

The standard candles most of interest are the globular clusters in
the halos of spiral galaxies: stellar evolution models lead to the
estimate~\cite{Peebles} that the age of oldest stars is $16\pm2
\times 10^9 \hbox{ yr}$.  The oldest stars seem to be forming
somewhat earlier than the development of galactic spiral structure
($ z_f \approx 15$).~\cite{Peebles} This bounds the Hubble
parameter~\cite{Galaxy,Zed15}

\begin{equation}
\hbox{SEC} \implies \quad 
H_0 \leq 58\pm7  \hbox{ km s$^{-1}$ Mpc$^{-1}$ }.
\end{equation}
Recent estimates of the present day value of the Hubble
parameter are~\cite{PDG}

\begin{equation} 
H_0 \in (65,85) \hbox{ km s$^{-1}$ Mpc$^{-1}$ }.
\end{equation}
But even the lowest reasonable value for the Hubble parameter is only
just barely compatible with the SEC, and that only by taking the
lowest reasonable value for the age of the globular clusters. For
currently favored values of the Hubble parameter we deduce that the
SEC is violated somewhere between the formation of the oldest stars
and the present time.~\cite{Galaxy,Zed15} {\em (Warning: observational
estimates of these quantities are changing rapidly as this is being
written.)}  If we reduce $z_f$ to be more in line with the formation
of the rest of the galactic structure (say $z_f\approx 7$) we make the
problem worse, not better ($H_0\leq54\pm7$ km s$^{-1}$
Mpc$^{-1}$). This difficulty is occurring at low cosmological
temperatures ($T \leq 60$ K), and late times, in a region where we
thought we understood the basic equation of state of the cosmological
fluid.

\section{SEC violations---Implications}

The two favorite ways of putting SEC violations into a classical field
theory are via a massive scalar field, or via a positive cosmological
constant.

A classical scalar field violates the SEC, but not the other energy
conditions. It is this violation of the SEC that makes cosmological
scalar fields so attractive to advocates of inflation.  Using a
massive scalar field to deal with the age-of-the-universe problem is
tantamount to asserting that a last dying gasp of inflation took place
as the galaxies were being formed. This would be extremely
surprising.~\cite{Galaxy,Zed15}

In contrast, the current favorite fix for the age-of-the-universe
problem is to introduce a positive cosmological constant.  Under the
mild constraint that the pressure due to normal matter in the present
epoch be non-negative, a SEC violation would require that more that
33\% of the present-day energy density is due to a cosmological
constant---and we do not need to know the equation of state of the
normal component of matter to extract this result.~\cite{Galaxy,Zed15}

\section{Discussion}

This analysis of the age-of-the-universe problem is, as far as
possible, model independent. I have shown that high values of the
Hubble parameter imply that the SEC must be violated sometime between
the epoch of galaxy formation and the present. This implies that the
age-of-the-universe problem cannot simply be fixed by adjusting the
equation of state of the cosmological fluid. Since all normal matter
satisfies the SEC, fixing the age-of-universe problem will inescapably
require the introduction of {\em ``abnormal''} matter---indeed we will
need large quantities of abnormal matter, sufficient to overwhelm the
gravitational effects of the normal matter.

\section*{Acknowledgments}

This research was supported by the U.S. Department of Energy.

\section*{References}

\end{document}